\documentclass[authoryear,preprint,12pt]{elsarticle}

\usepackage{amssymb}
\usepackage{amsmath}
\usepackage{graphicx}
\usepackage{booktabs}
\usepackage{array}
\usepackage{url}

\journal{Journal Name}

\begin{document}

\begin{frontmatter}

\title{Global Trends and Heterogeneity in Modified Adjusted Net Savings (1990–2021)}

\author[1]{Xin Xiao\corref{cor1}}
\ead{xiaoxin@stiei.edu.cn}
\cortext[cor1]{Corresponding author}

\affiliation[1]{organization={School of Electronic Technology and Engineering, Shanghai Technical Institute of Electronics and Information},
            addressline={Street Address},
            city={City},
            postcode={000000},
            state={State},
            country={Country}}

\begin{abstract}
Modified adjusted net savings (MANS) is a derived weak-sustainability indicator obtained by subtracting education expenditure from adjusted net savings (ANS). Because ANS equals net national savings plus education expenditure minus natural-resource depletion and carbon dioxide damage, MANS isolates the savings and depletion components and is useful when education expenditure would otherwise dominate cross-country comparisons. This study documents the global distribution, long-run trends, and cross-country heterogeneity of MANS excluding particulate emission damage over 1990--2021. The accounting base is the World Bank World Development Indicators series NY.ADJ.SVNX.CD, and MANS is constructed by subtracting education expenditure from this series. We combine global maps, aggregate time series, country-level heatmaps, distributional statistics, and income-group comparisons. The results show that global MANS has increased substantially in aggregate, but its evolution as a share of nominal GDP has been more volatile and uneven across countries. Per-capita MANS and global MANS totals remain concentrated in high-income and large economies, whereas low-income and lower-middle-income groups display persistently lower and more volatile outcomes. Country rankings are highly persistent, and some populous countries exhibit pronounced fluctuations associated with natural-resource dependence and structural change. The study provides a reproducible descriptive benchmark for monitoring sustainable development and highlights the need to interpret MANS in current US dollars alongside income-group and country-specific contexts. The findings are relevant for macroeconomic policy, natural-capital accounting, and the design of sustainability-oriented investment strategies.
\end{abstract}


\begin{highlights}
\item Global modified adjusted net savings (MANS) is documented for 1990--2021.
\item Aggregate MANS rises, but MANS/GDP remains volatile across countries and years.
\item Per-capita MANS and totals concentrate in high-income and large economies.
\item Low-income groups show persistently lower and more volatile MANS outcomes.
\item Country ranks persist; resource dependence shapes within-country swings.
\end{highlights}


\end{frontmatter}

\section{Introduction}

Measuring whether economies are on a sustainable development path requires accounting for changes in produced, human, and natural capital. Adjusted net savings (ANS), also referred to as genuine savings, extends conventional net national savings by adding education expenditure and subtracting energy depletion, mineral depletion, net forest depletion, and carbon dioxide damage \citep{pearce1993,hamilton1999}. Under a weak-sustainability interpretation, a positive ANS indicates that an economy is accumulating total wealth after accounting for natural-resource depletion and environmental damage, whereas persistently negative ANS raises concerns about the erosion of the productive base \citep{dasgupta2001,arrow2004}. The theoretical foundation of this interpretation rests on the assumption that different forms of capital are substitutable, an assumption that remains debated but is central to the policy use of ANS \citep{cohen2019,hamilton2014,dietz2007,neumayer2013}.

The World Bank has produced and maintained ANS estimates for a large panel of economies, and these estimates are now part of the World Development Indicators (WDI) \citep{worldbank2024a}. The accounting base used here is the WDI series NY.ADJ.SVNX.CD: adjusted net savings excluding particulate emission damage, expressed in current US dollars. The indicator equals net national savings plus education expenditure minus energy depletion, mineral depletion, net forest depletion, and carbon dioxide damage. Particulate emission damage is excluded from this particular series. Because the indicator is expressed in current prices, it should be interpreted as a current-dollar measure rather than a constant-price real series \citep{worldbank2024b}. Recent methodological work argues that the World Bank ANS measure could be further refined by accounting for productivity growth, population change, and undervalued human-capital investment \citep{pezzey2024}; we return to this caveat in the discussion.

This paper focuses on modified adjusted net savings (MANS), defined as ANS minus education expenditure. The transformation isolates the savings, resource-depletion, and carbon-damage components of ANS and avoids conflating cross-country comparisons with differences in education expenditure. MANS is therefore the common analytical object in all figures and tables. ANS is retained only as the accounting base from which MANS is derived. The construction follows the adjusted-net-savings methodology summarized by \citet{bolt2002}.

Despite the widespread use of ANS and related genuine-savings measures in sustainability monitoring, the global evidence is often presented in isolated country studies or aggregate tables. A systematic descriptive account of global distribution, long-run trends, and income-group heterogeneity in MANS is useful for at least three reasons. First, MANS is a composite indicator whose components can move in opposite directions, so aggregate trends may conceal important cross-country differences. Second, current-dollar MANS totals are strongly influenced by economy size, whereas per-capita and GDP-share measures provide complementary information. Third, the persistence of country rankings and the volatility of resource-dependent economies are central to understanding whether observed improvements are broad-based or concentrated. The resource-curse literature shows that natural-resource rents and exports can lower genuine savings in many countries, with institutional quality shaping the outcome \citep{vankrevel2024}; this makes the heterogeneity analysis especially relevant.

This paper contributes a reproducible descriptive benchmark for global MANS excluding particulate emission damage over 1990--2021. We organize the evidence around four questions: (i) How has global MANS evolved in aggregate and as a share of nominal GDP? (ii) How unequal is MANS per capita across countries and regions? (iii) How persistent are country rankings and how heterogeneous are populous countries? (iv) How do MANS outcomes differ across income groups? We use global maps, aggregate time series, country-level heatmaps, distributional statistics, and income-group comparisons. The analysis is descriptive and exploratory; it does not claim causal identification. Instead, it provides stylized facts that can guide more structural work on natural-capital accounting and sustainability policy.

The remainder of the paper is organized as follows. Section~2 describes the data and the MANS indicator. Section~3 presents the empirical framework. Section~4 reports the results. Section~5 discusses implications and limitations, and Section~6 concludes.

\section{Data and indicator}

\subsection{Data source}

The analysis uses the World Bank WDI series NY.ADJ.SVNX.CD, adjusted net savings excluding particulate emission damage, for the period 1990--2021. The series is expressed in current US dollars. The underlying national-accounts aggregates are compiled according to international standards, including the System of National Accounts 2008 or 1993 versions, and the ANS estimates are produced by World Bank staff estimates \citep{worldbank2024a,worldbank2021}. We use the annual country-level panel as provided in WDI. The sample is unbalanced because data availability differs across countries and years.

The accounting base is defined as
\begin{equation}
\mathrm{ANS}_{it} = \mathrm{NNS}_{it} + \mathrm{EDU}_{it} - \mathrm{ED}_{it} - \mathrm{MD}_{it} - \mathrm{NFD}_{it} - \mathrm{CO}_{2,it},
\label{eq:ans}
\end{equation}
where $\mathrm{NNS}_{it}$ is net national savings, $\mathrm{EDU}_{it}$ is education expenditure, $\mathrm{ED}_{it}$ is energy depletion, $\mathrm{MD}_{it}$ is mineral depletion, $\mathrm{NFD}_{it}$ is net forest depletion, and $\mathrm{CO}_{2,it}$ is carbon dioxide damage. The series used here excludes particulate emission damage. All components are measured in current US dollars.

The analytical indicator used throughout the paper is MANS, defined as
\begin{equation}
\mathrm{MANS}_{it} = \mathrm{ANS}_{it} - \mathrm{EDU}_{it}
= \mathrm{NNS}_{it} - \mathrm{ED}_{it} - \mathrm{MD}_{it} - \mathrm{NFD}_{it} - \mathrm{CO}_{2,it}.
\label{eq:mans}
\end{equation}
Equation~\eqref{eq:mans} shows that subtracting $\mathrm{EDU}_{it}$ exactly cancels the $+\mathrm{EDU}_{it}$ term in Eq.~\eqref{eq:ans}; education expenditure therefore does not appear in the final MANS expression. MANS is not the headline WDI series and should not be conflated with NY.ADJ.SVNX.CD. The headline WDI ANS series is used only as the accounting base for constructing MANS.

\subsection{Interpretation and measurement caveats}

MANS inherits the weak-sustainability interpretation of ANS. It assumes that different forms of capital are substitutable and that a non-declining total capital stock is consistent with sustainable development. This interpretation is useful for macroeconomic monitoring but should not be treated as a comprehensive measure of ecological sustainability \citep{cohen2019,dietz2007,neumayer2013}. The valuation of natural-resource depletion and environmental damage depends on prices, rents, and country-specific accounting practices. Moreover, current-dollar values are not adjusted for inflation or exchange-rate movements over time. For these reasons, we report MANS both in levels and as a share of nominal GDP where appropriate, and we complement global aggregates with per-capita and distributional measures.

Table~\ref{tab:components} summarizes the components, their signs in Eqs.~\eqref{eq:ans} and \eqref{eq:mans}, and their treatment in MANS.

\begin{table}[t]
\centering
\caption{Components and interpretation of ANS and MANS excluding particulate emission damage.}
\label{tab:components}
\footnotesize
\setlength{\tabcolsep}{3pt}
\begin{tabular}{@{}p{0.17\linewidth}c p{0.22\linewidth}c p{0.31\linewidth}@{}}
\toprule
Component & Sign in Eq.~\eqref{eq:ans} & Treatment in MANS & Sign in Eq.~\eqref{eq:mans} & Interpretation \\
\midrule
Net national savings & $+$ & Retained & $+$ & Savings after depreciation of produced capital. \\
Education expenditure & $+$ & Removed by subtracting $\mathrm{EDU}_{it}$ & $0$ & Investment in human capital; cancels the $+\mathrm{EDU}_{it}$ term and is absent from MANS. \\
Energy depletion & $-$ & Retained & $-$ & Depletion of energy resources valued at resource rents. \\
Mineral depletion & $-$ & Retained & $-$ & Depletion of mineral resources valued at resource rents. \\
Net forest depletion & $-$ & Retained & $-$ & Net depletion of forest resources. \\
Carbon dioxide damage & $-$ & Retained & $-$ & Damage from CO$_2$ emissions. \\
Particulate emission damage & Not included & Not included & Not included & This series excludes particulate emission damage. \\
\bottomrule
\end{tabular}

\par\smallskip
\begin{minipage}{0.98\linewidth}
\footnotesize\raggedright
Note: NY.ADJ.SVNX.CD excludes particulate emission damage, so this component is absent from both Eq.~\eqref{eq:ans} and Eq.~\eqref{eq:mans}. The $+$ and $-$ signs refer to the component's contribution in the corresponding equation; $0$ means the term is cancelled and does not appear in the final MANS expression.
\end{minipage}
\end{table}

\section{Empirical framework}

The empirical analysis is descriptive and uses several complementary representations of the same underlying panel.

First, we map the global distribution of MANS per capita excluding particulate emission damage and education expenditure for benchmark years 1990, 2000, 2010, and 2021. The maps are used to identify broad spatial patterns and changes in the cross-country distribution. The color scale reports ranks in percent, which facilitates comparison across years despite changes in the level of the indicator.

Second, we aggregate the country-level MANS series to the global level and compare global MANS with global GDP. We report MANS in billions of current US dollars and MANS as a percentage of nominal GDP. This distinguishes changes in the level of MANS from changes in its relative importance in the world economy.

Third, we examine the sample total, the median country, and the interquartile range of MANS. The sample total captures the behavior of large economies, whereas the median and interquartile range summarize the typical country and cross-country dispersion. We also report the cross-country standard deviation and the P90--P10 range to describe convergence or divergence.

Fourth, we visualize the top 30 most populous countries in 2021, sorted by their 2021 MANS per capita. This heatmap highlights within-country dynamics and the persistence of country rankings. We complement it with a ranking of average MANS per country in billions of current US dollars (total, not per capita), which is used only to illustrate the dominance of large economies and is not directly comparable to the per-capita heatmap.

Fifth, we compare MANS across income groups. We report medians and interquartile ranges for high-income, upper-middle-income, lower-middle-income, and low-income groups. We also report the contribution of each income group to global MANS in 2021. These comparisons are descriptive and should not be interpreted as causal effects of income status.

Finally, we present exploratory heterogeneity analyses by income group and region. The fixed-effect coefficients reported in the supplementary figures are descriptive conditional associations, not causal estimates. The goal is to identify robust stylized facts that can motivate further research.

\section{Results}

\subsection{Global distribution of MANS per capita}

\begin{figure}[t]
\centering
\includegraphics[width=\linewidth]{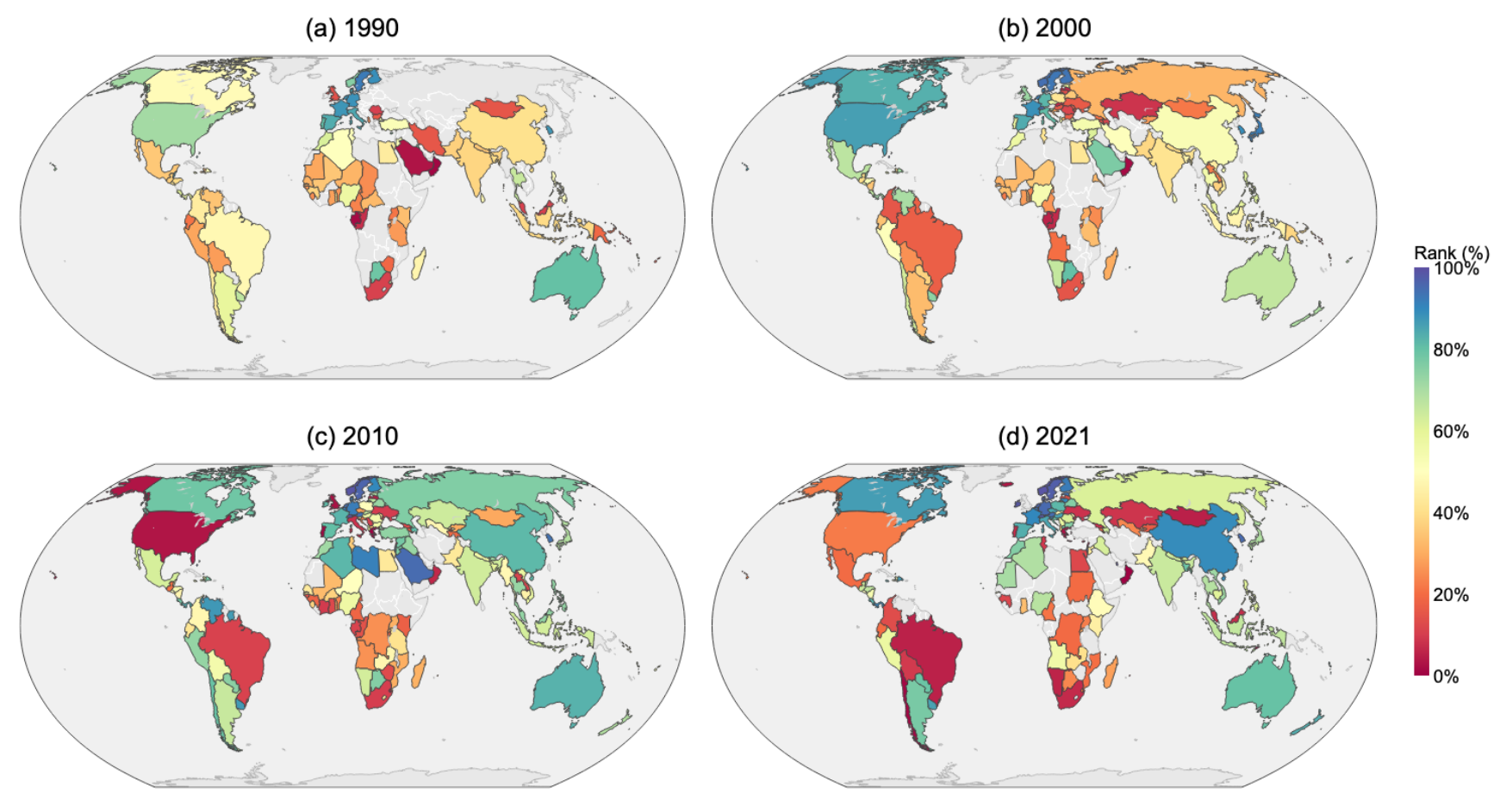}
\caption{Global distribution of modified adjusted net savings (MANS) per capita, 1990--2021. MANS is ANS minus education expenditure. Colour scale reports percentile ranks.}
\label{fig:map}
\end{figure}

Figure~\ref{fig:map} reports the global distribution of MANS per capita for 1990, 2000, 2010, and 2021.

Three patterns stand out. First, high-income countries in North America, Western Europe, and parts of East Asia tend to occupy the upper part of the distribution. Second, resource-rich economies in Latin America, Central Asia, and Sub-Saharan Africa display more heterogeneous and volatile positions, consistent with the sensitivity of MANS to energy and mineral depletion and with recent causal evidence that energy rents and exports can lower genuine savings in the absence of strong institutions \citep{vankrevel2024}. Third, the cross-country distribution has not converged uniformly: some countries improve their relative position, whereas others experience declines or persistent low ranks. The maps therefore suggest that global progress in MANS is not evenly shared.

\subsection{Aggregate trends and GDP share}

\begin{figure}[t]
\centering
\includegraphics[width=0.5\linewidth]{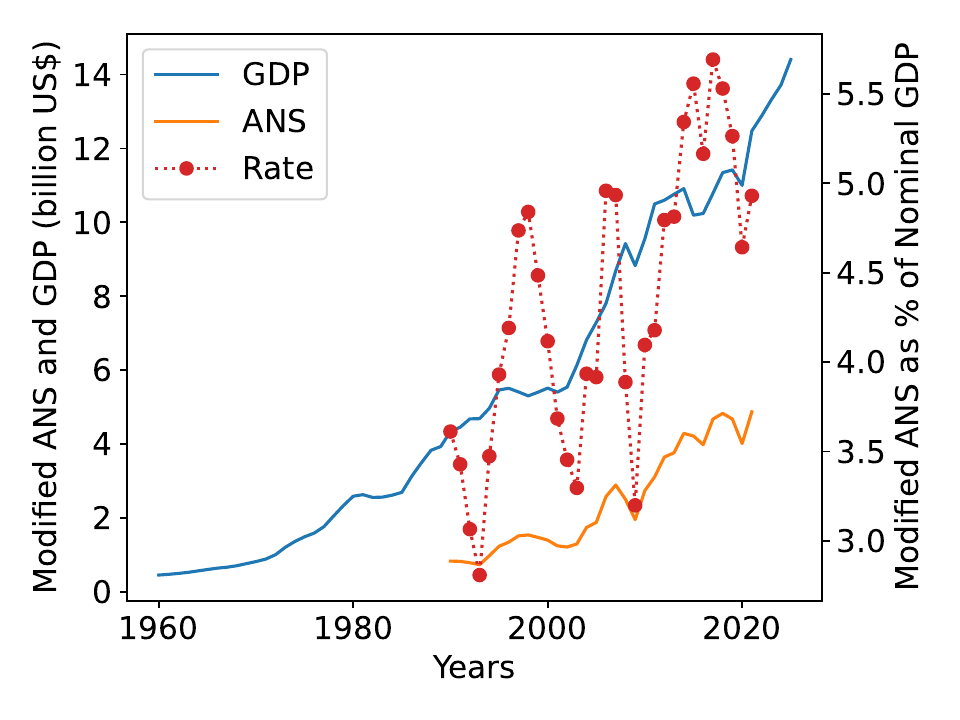}
\caption{Global modified adjusted net savings (MANS) and MANS as a percentage of nominal GDP, 1960--2021. The analysis window is 1990--2021.}
\label{fig:world}
\end{figure}

Figure~\ref{fig:world} shows global MANS and MANS as a percentage of nominal GDP. The underlying series spans 1960--2021; the 1990--2021 window is the focus of this paper, while the pre-1990 segment is retained as background context.

Global MANS in billions of current US dollars rises substantially over the period, with particularly rapid growth after the early 2000s. However, MANS as a share of nominal GDP is much more volatile. The gap between the level and the share indicates that part of the increase in current-dollar MANS reflects the growth of the world economy and price effects rather than a proportional improvement in savings behavior. The volatility of the GDP share also suggests that global MANS is sensitive to commodity cycles, macroeconomic fluctuations, and changes in natural-resource rents.

\subsection{Sample totals, medians, and dispersion}

\begin{figure}[t]
\centering
\includegraphics[width=\linewidth]{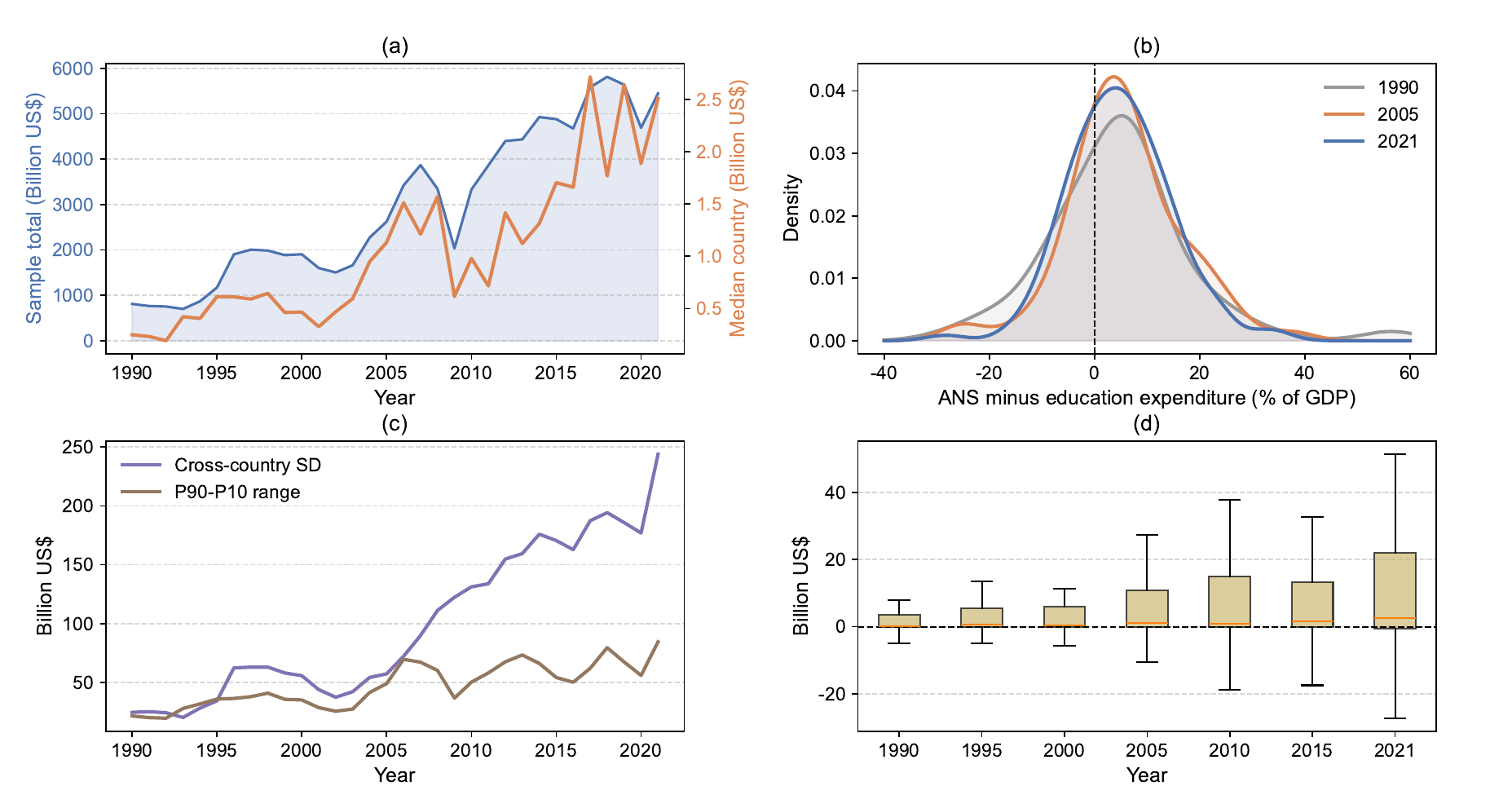}
\caption{Distributional summaries of modified adjusted net savings (MANS), 1990--2021. Panel (a): sample total; panel (b): kernel density of MANS/GDP for 1990, 2005 and 2021; panel (c): cross-country standard deviation and P90--P10 range; panel (d): distribution of MANS/GDP across countries over time.}
\label{fig:stat}
\end{figure}

Figure~\ref{fig:stat} reports four complementary summaries of MANS over 1990--2021: (a) the sample total, (b) the cross-country distribution for selected years, (c) the cross-country standard deviation and the P90--P10 range, and (d) the distribution of MANS as a share of GDP.

Panel~(a) shows that the sample total of MANS increases strongly, with noticeable fluctuations around major global shocks. Panel~(b) shows that the distribution of MANS as a share of GDP has shifted over time while retaining a long right tail. Panel~(c) reports the cross-country standard deviation and the P90--P10 range; both dispersion measures indicate that cross-country differences are substantial and, in some periods, widening. Panel~(d) confirms that the distribution is centered near zero but has a long right tail, implying that a subset of countries achieves very high MANS shares while others remain close to zero or negative.

\subsection{Populous countries and country rankings}

\begin{figure}[t]
\centering
\includegraphics[width=\linewidth]{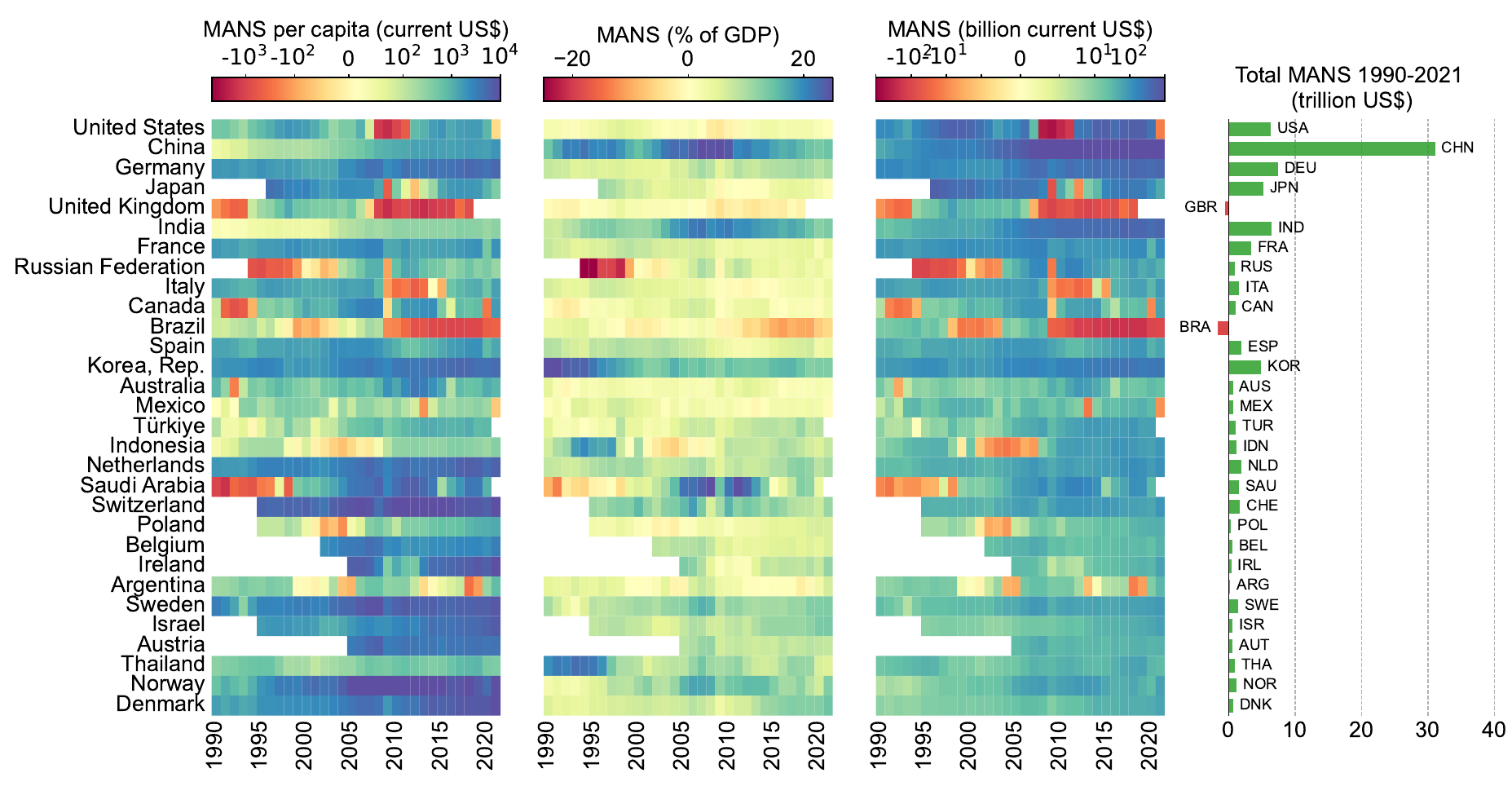}
\caption{Modified adjusted net savings (MANS) per capita for the top 30 most populous countries in 2021, sorted by 2021 values in descending order. Left column: MANS per capita (current US\$); middle column: MANS as a share of GDP; right column: total MANS (billion current US\$).}
\label{fig:heat}
\end{figure}

Figure~\ref{fig:heat} presents MANS per capita for the top 30 most populous countries in 2021, sorted by their 2021 values in descending order.

The heatmap reveals strong persistence in country rankings: countries with high per-capita MANS in the early 1990s tend to remain high, and countries with low values tend to remain low. At the same time, several populous countries exhibit pronounced within-country changes. These fluctuations are consistent with the combined influence of savings rates, resource rents, and carbon dioxide damage, and with the finding that resource-dependent economies can experience large swings in genuine savings \citep{vankrevel2024}. The results also show that population size and per-capita MANS are distinct dimensions: some populous countries have high totals but modest per-capita values, whereas others have high per-capita values but smaller global totals.

\begin{figure}[t]
\centering
\includegraphics[width=\linewidth]{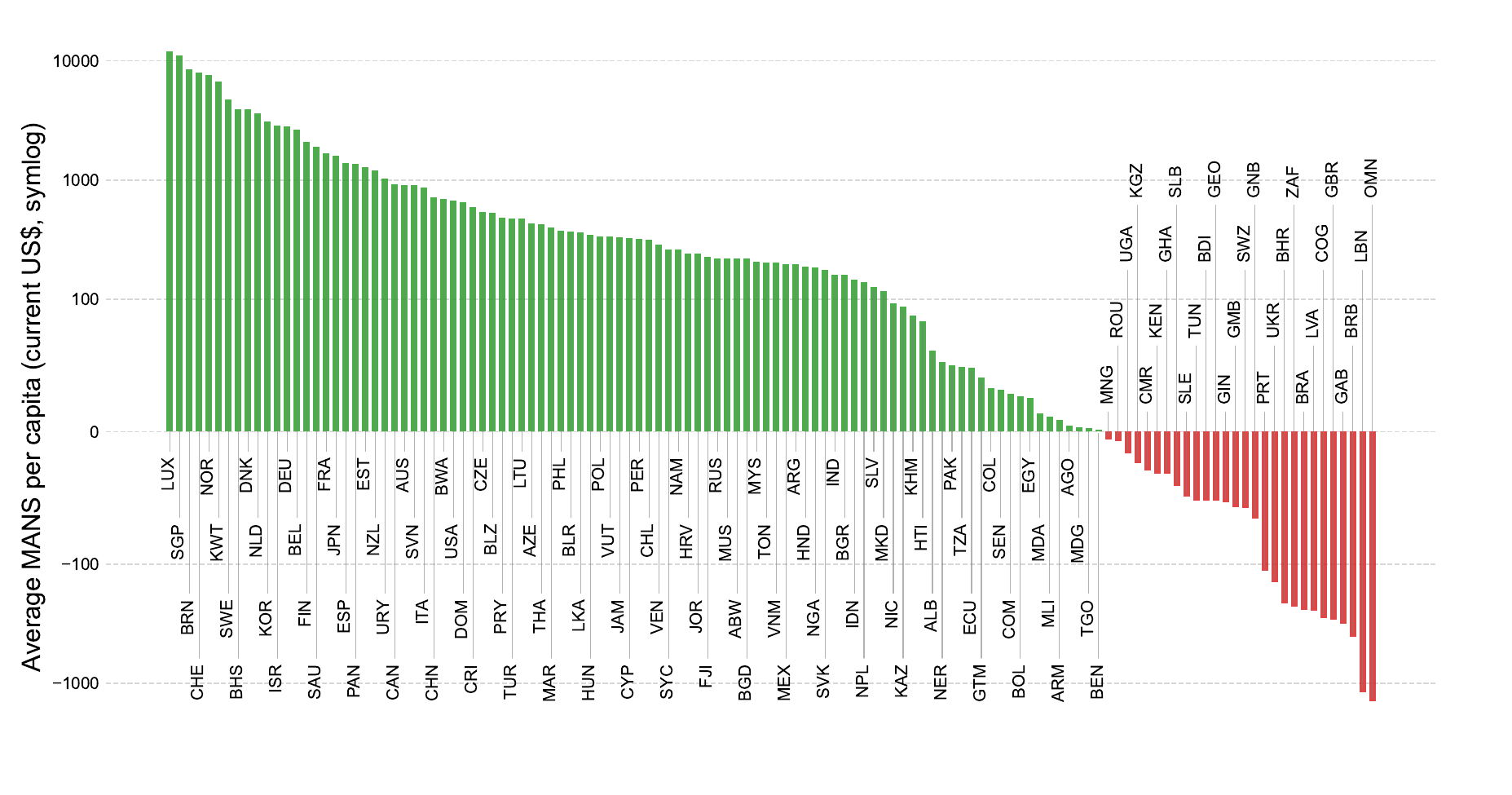}
\caption{Average modified adjusted net savings (MANS) per capita by country, ranked in current US dollars (symlog scale). This figure is not a ranking of total MANS in billions.}
\label{fig:mean}
\end{figure}

Figure~\ref{fig:mean} ranks countries by average MANS per capita in current US dollars, using a symmetric logarithmic scale to accommodate both positive and negative values.

The ranking is dominated by a mix of high-income and resource-rich economies at the top and by low-income and resource-dependent economies at the bottom. Because the vertical axis is per capita, this figure should not be read as a ranking of total contributions to global MANS; total contributions are mechanically dominated by large economies, as noted above. The contrast between per-capita and total rankings reinforces the need to interpret MANS alongside GDP-share measures.

\subsection{Income-group heterogeneity}

\begin{figure}[t]
\centering
\includegraphics[width=\linewidth]{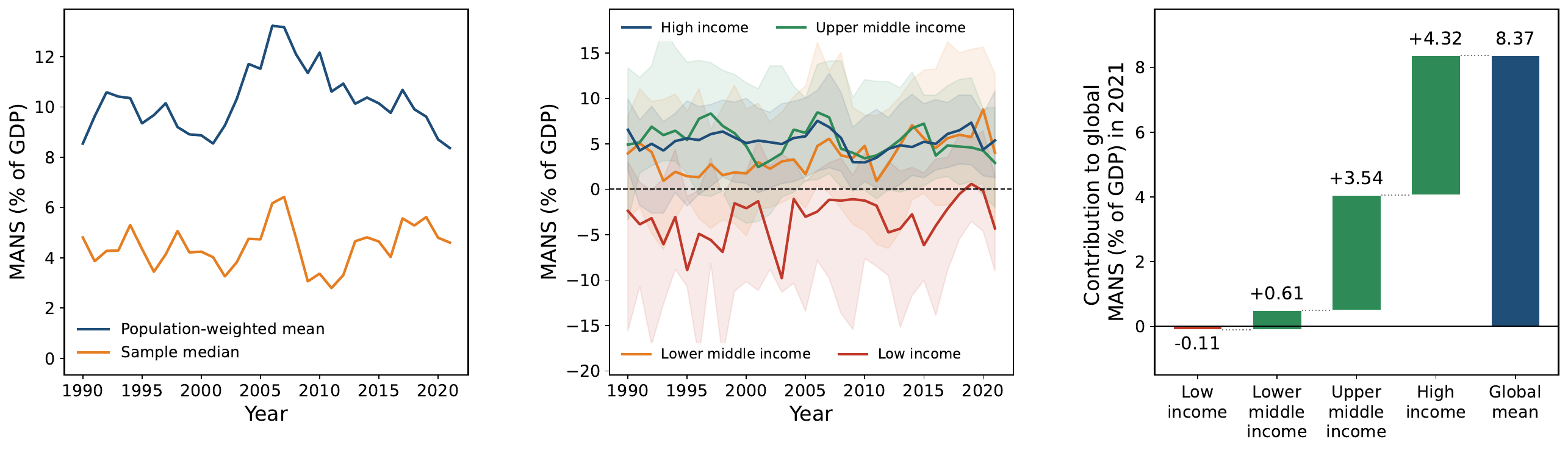}
\caption{Modified adjusted net savings (MANS) by income group, 1990--2021. Left panel: population-weighted mean and sample median; middle panel: medians with interquartile ranges by income group; right panel: contribution of each income group to global MANS in 2021 (\% of global GDP).}
\label{fig:income}
\end{figure}

Figure~\ref{fig:income} compares MANS across income groups.

High-income countries report the highest median MANS as a share of GDP, followed by upper-middle-income countries. Lower-middle-income and low-income groups display lower medians and wider interquartile ranges. The gap between high-income and low-income groups is persistent over the sample period. Figure~\ref{fig:income} also shows that the contribution of income groups to global MANS in 2021 is highly concentrated: high-income countries account for the largest share, while low-income countries contribute a very small share. These patterns are consistent with the concentration of economic size and savings capacity, but they also raise questions about the sustainability of development paths in lower-income economies.

\begin{figure}[t]
\centering
\includegraphics[width=\linewidth]{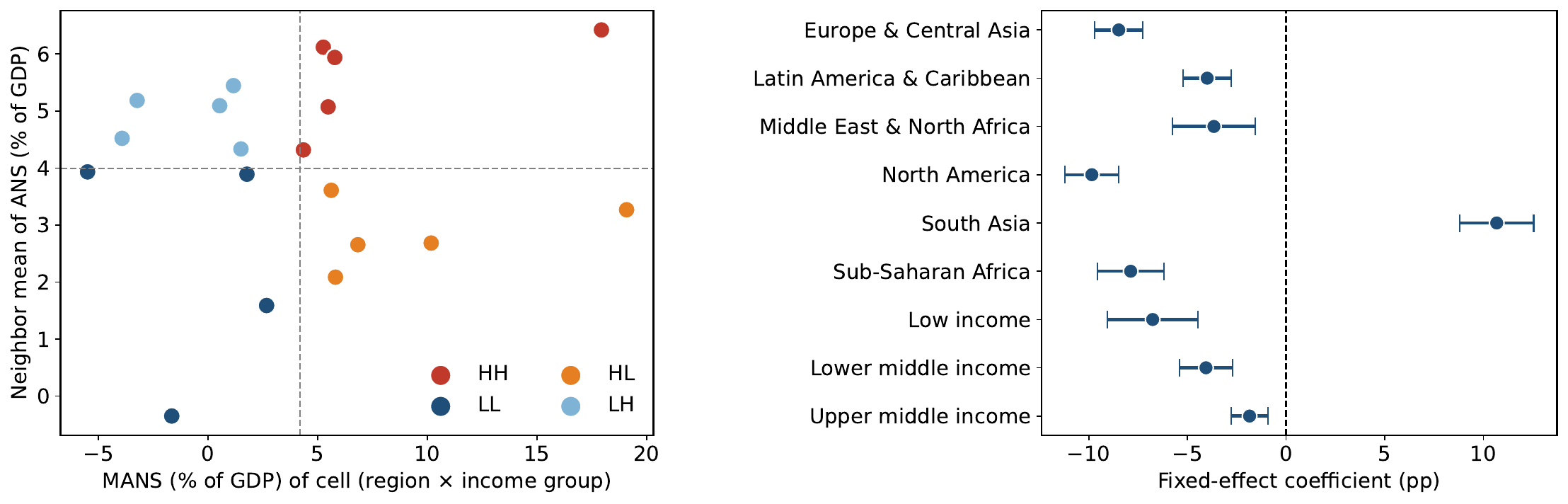}
\caption{Exploratory income-group and regional heterogeneity in modified adjusted net savings (MANS). Left panel: mean MANS by income-group cell; right panel: fixed-effect coefficients with confidence intervals. These are descriptive conditional associations, not causal estimates.}
\label{fig:lisa}
\end{figure}

Figure~\ref{fig:lisa} provides an exploratory heterogeneity analysis by income group and region.

The left panel shows the association between MANS as a share of GDP and income-group cells, while the right panel reports fixed-effect coefficients with confidence intervals. The coefficients should be read as descriptive conditional associations. The results indicate that regional and income-group differences remain after accounting for common time effects, although the precision of the estimates varies across groups.

\begin{figure}[t]
\centering
\includegraphics[width=\linewidth]{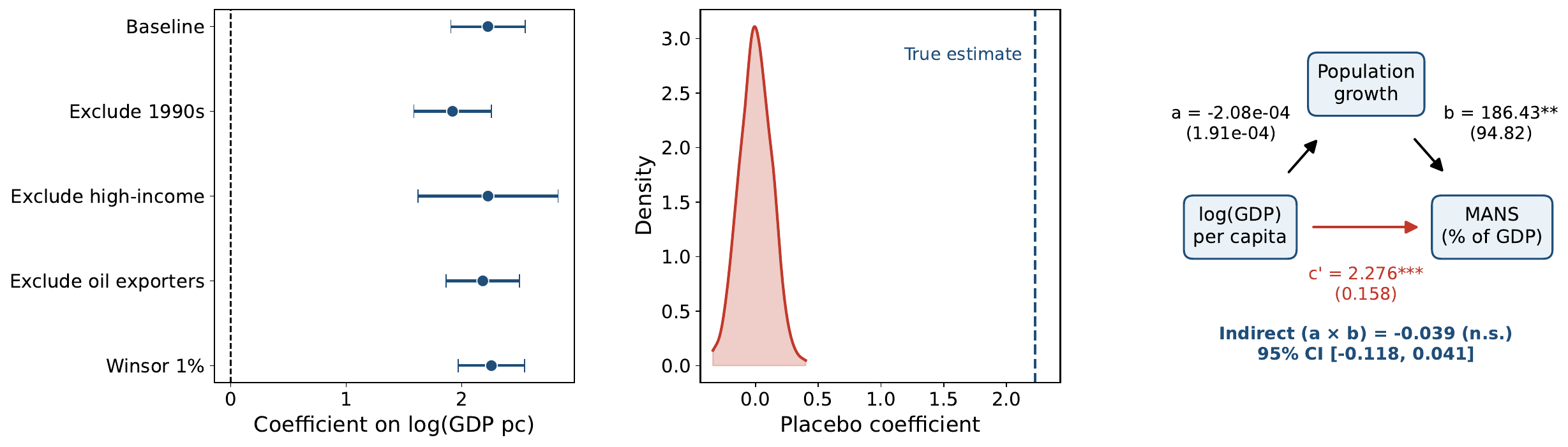}
\caption{Additional robustness checks for modified adjusted net savings (MANS) by income group, 1990--2021.}
\label{fig:stable}
\end{figure}

Figure~\ref{fig:stable} reports additional robustness checks for MANS by income group.

The qualitative ranking of income groups is stable across alternative summaries, although the level and dispersion of MANS vary with the chosen measure. This stability supports the interpretation that income-group differences are a robust feature of the data rather than an artifact of a single aggregation method.

\section{Discussion}

The descriptive evidence in this paper yields four main insights. First, global MANS has grown substantially in current US dollars, but MANS as a share of nominal GDP has not followed a smooth upward path. This distinction is important for policy: a rising current-dollar total may reflect economic growth and price effects, whereas a volatile GDP share points to structural and cyclical pressures on savings and natural-capital depletion. Second, the global distribution of MANS per capita is uneven and persistent. High-income countries and some resource-rich economies occupy the upper ranks, while many low-income and lower-middle-income countries remain in the lower part of the distribution. Third, country rankings are highly persistent, but several populous countries experience large within-country swings. This suggests that country-specific factors, including resource dependence and structural transformation, matter for sustainability outcomes, consistent with causal evidence that energy rents and exports lower genuine savings in many countries unless institutional safeguards are in place \citep{vankrevel2024}. Fourth, income-group gaps are robust across alternative summaries, with high-income countries contributing the largest share of global MANS and low-income countries contributing very little.

These findings have implications for sustainability monitoring. MANS is useful because it integrates savings, resource depletion, and carbon damage into a single accounting framework while removing the education component that can dominate cross-country comparisons. However, it should not be used as a standalone measure of sustainability. Current-dollar MANS is sensitive to price and exchange-rate movements, and the weak-sustainability assumption of capital substitutability is strong \citep{cohen2019,dietz2007,neumayer2013}. Recent methodological work further suggests that the World Bank ANS could be refined by accounting for productivity growth, population change, and human-capital depreciation \citep{pezzey2024}; the stylized facts reported here should therefore be read as a benchmark for the existing WDI-derived MANS series rather than as a definitive statement on sustainability. Policymakers should combine MANS with physical indicators of natural-capital stocks, emissions, biodiversity, and distributional measures \citep{yamaguchi2023}. The income-group gaps documented here also suggest that international support for human-capital investment and natural-capital accounting may be important for lower-income economies.

Several limitations should be acknowledged. The analysis is descriptive and does not identify causal mechanisms. The WDI series is unbalanced, and country-level accounting practices may differ. The valuation of natural-resource depletion and carbon damage involves assumptions that are not uniform across countries. The exploratory fixed-effect coefficients should not be interpreted as causal effects. Finally, because MANS subtracts education expenditure, it isolates a particular component of ANS and should not be confused with the headline WDI series NY.ADJ.SVNX.CD. Future research could link the stylized facts documented here to structural models of savings, resource rents, and environmental policy.

\section{Conclusion}

This paper provides a reproducible descriptive benchmark for global modified adjusted net savings excluding particulate emission damage over 1990--2021. Using World Bank WDI data and constructing MANS by subtracting education expenditure from the ANS series NY.ADJ.SVNX.CD, we document rising global MANS in current US dollars, volatile MANS as a share of nominal GDP, persistent cross-country inequality, strong country-rank persistence, and substantial income-group gaps. The evidence highlights that aggregate progress in MANS masks considerable heterogeneity across countries and income groups. For high-level sustainability monitoring, MANS should be interpreted alongside per-capita measures, GDP shares, natural-capital stock indicators, and country-specific context. The figures and stylized facts reported here can serve as a starting point for more structural research on sustainable development and natural-capital accounting.

\section*{Declaration of competing interest}

The authors declare that they have no known competing financial interests or personal relationships that could have appeared to influence the work reported in this paper.

\section*{Data availability}

The data used in this study are publicly available from the World Bank World Development Indicators. The accounting-base indicator is NY.ADJ.SVNX.CD; MANS is derived by subtracting education expenditure from this series. Derived figures and code can be made available from the corresponding author upon reasonable request.

\section*{Acknowledgements}

This research received no specific grant from any funding agency in the public, commercial, or not-for-profit sectors. The authors thank the World Bank for maintaining the World Development Indicators database. Any remaining errors are our own.

\section*{CRediT authorship contribution statement}

\textbf{First Author:} Conceptualization, Data curation, Formal analysis, Methodology, Writing -- original draft. \textbf{Second Author:} Validation, Visualization, Writing -- review and editing.


\end{document}